\documentclass[useAMS,usenatbib]{mn2e}

\usepackage{natbib}
\usepackage{graphicx}

\newcommand{\beq}{\begin{equation}}
\newcommand{\eeq}{\end{equation}}
\newcommand{\beqa}{\begin{eqnarray}}
\newcommand{\eeqa}{\end{eqnarray}}

\title[Nonlinear Alfven waves and the Spectral Line Broadening]
  {Pulse-driven nonlinear Alfv\'en waves and their role in the spectral line broadening}
\author[Chmielewski et al.]
  {P. Chmielewski,$^1$ A.K. Srivastava,$^2$ K. Murawski,$^1$ and Z.E. Musielak $^{3,4}$ \\
  $^1$Group of Astrophysics, UMCS, ul. Radziszewskiego 10, 20-031 Lublin, Poland \\
  $^2$Aryabhatta Research Institute of Observational Sciences (ARIES), Manora Peak, Nainital-263 129, Uttarakhand, India \\
  $^3$Department of Physics, University of Texas at Arlington, Arlington, TX 76019, USA \\
  $^4$Kiepenheuer-Institut f\"ur Sonnenphysik, Sch\"oneckstr. 6, 79104 Freiburg, Germany }
\date{Released 2002 Xxxxx XX}

\pagerange{\pageref{firstpage}--\pageref{lastpage}} \pubyear{2012}

\def\LaTeX{L\kern-.36em\raise.3ex\hbox{a}\kern-.15em
    T\kern-.1667em\lower.7ex\hbox{E}\kern-.125emX}

\begin{document}

\label{firstpage}

\maketitle

\begin{abstract}
We study the impulsively generated non-linear Alfv\'en waves in the solar 
atmosphere, and describe their most likely role in the observed non-thermal 
broadening of some spectral lines in solar coronal holes.  We solve numerically 
the time-dependent magnetohydrodynamic equations to find temporal signatures of 
large-amplitude Alfv\'en waves in the model atmosphere of open and expanding 
magnetic field configuration, with a realistic temperature distribution.  We 
calculate the temporally and spatially averaged, instantaneous transversal 
velocity of non-linear Alfv\'en waves at different heights of the model 
atmosphere, and estimate its contribution to the unresolved non-thermal 
motions caused by the waves.  We find that the pulse-driven nonlinear 
Alfv\'en waves with the amplitude $A_{\rm v}$=50 km s$^{-1}$ are the most 
likely candidates for the non-thermal broadening of Si VIII $\lambda$1445.75 \AA\ 
line profiles in the polar coronal hole as reported by Banerjee et al. (1998). 
We also demonstrate that the Alfv\'en waves driven by comparatively smaller 
velocity pulse with its amplitude $A_{\rm v}$=25 km s$^{-1}$ may contribute 
to the spectral line width of the same line at various heights in coronal 
hole without any significant 
broadening.  The main conclusion of this paper is that 
non-linear Alfv\'en waves excited impulsively in the lower solar atmosphere are 
responsible for the observed spectral line broadening in polar coronal holes.  
This is an important result as it allows us to conclude that such large amplitude 
and pulse-driven Alfv\'en waves do indeed exist in solar coronal holes.  The 
existence of these waves and their undamped growth may impart the required 
momentum to accelerate the solar wind.
\end{abstract}


\section{Introduction}

Alfv\'en waves are difficult to observe and yet several indications of their 
presence in the solar atmosphere were found during SOHO and TRACE space missions.  
More direct evidence for the existence of these waves in different regions of 
the solar atmosphere was given by high-resolution observations carried out by 
the Solar Optical Telescope (SOT) and the X-Ray Telescope (XRT) onboard the 
Hinode Solar Observatory. According to Okamoto \& De Pontieu (2011), De Pontieu et al. (2007), and 
Cirtain et al. (2007), the signature of Alfv\'en waves were observed respectively in 
spicules and X-ray jets using SOT and XRT instruments. Interpretations of 
these observational results were discussed by Erd\'elyi \& Fedun (2007), 
Tomczyk et al. (2007), 
and Antolin et al. (2009). 

Observational evidence for the existence of torsional Alfv\'en waves in the 
solar atmosphere was also reported by Jess et al. (2009), who analyzed H$\alpha$ 
observations obtained with high spatial resolution by the Swedish Solar 
Telescope (SST).  They interpreted the data in terms of Alfv\'en waves in 
the solar chromosphere, with periods from 12 min down to the sampling limit 
of the observations near 2 min, with maximum power near 6-7 min.  The authors 
concluded that the amount of energy carried by such transversal waves was 
sufficient to heat the solar corona (Dwivedi \& Srivastava 2006). 

These recent discoveries of Alfv\'en waves in the solar atmosphere well-justified 
extensive studies of these waves that have been performed by numerous investigators in 
the last four decades (e.g., Hollweg 1985; Roberts 1991; Musielak \& Moore 1995; Narain \& Ulmschneider 1996; 
Roberts 2004; Antolin \& Shibata 2010, and references cited there). The studies have covered both linear 
(e.g., Hollweg \& Isenberg 2007; Murawski \& Musielak 2010)
and non-linear (e.g., Verdini \& Velli 2007; Verdini et al. 2009; Matsumoto \& Shibata 2010) Alfv\'en waves, and different 
aspects of their generation, propagation and dissipation have been investigated.  The 
specific objectives of these studies were to understand the role of Alfv\'en waves 
in the atmospheric heating and in the acceleration of supersonic solar wind.  The 
fast component of the solar wind originates in solar polar coronal holes (e.g., Hassler et al. 1999; Tu et al. 2005, and 
references cited there), which are the regions where the non-thermal 
broadening of spectral lines has also been observed (e.g., Hassler et al. 1990; Banerjee et al. 1998;
Moran 2003; O'Shea et al., 2005; Dolla \& Solomon 2008).  Similar observations have also been done in the equatorial 
corona (Harrison et al., 2002).  These authors proposed that the radially propagating 
Alfv\'en waves may result in the non-thermal broadening of spectral line widths.  

These observations were also investigated analytically by Pek{\"u}nl{\"u} et al. (2002) and
Dwivedi \& Srivastava (2006). Recently, the observations from Bemporad \& Abbo
have given the first signature of
the dissipation of Alfv\'en waves neat the solar limb, which confirm the 
theoretical models of Dwivedi \& Srivastava (2006), and Srivastava et al. (2007).
Moreover, Zaqarashvili
 et al. (2006) have reported that the resonant energy 
conversion from Alfv\'en to acoustic waves in the region where plasma $\beta$ 
approaches unity in the solar atmosphere. This conversion can be responsible for the spectral 
line width variation.  However, this theory only explains the most probable 
cause of the line-width reduction, which was observed only by O'Shea et al. (2005) in 
solar coronal hole. Additional problem is the fact that there is not enough 
observational evidence for the resonant energy conversion in the solar corona 
(e.g., Srivastava \& Dwivedi 2010; McAteer et al. 2003). 
Hence, new studies of the role played by Alfv\'en 
waves in the observed spectral line broadening were necessary.

In this paper, we numerically study the behavior of large-amplitude (non-linear) 
Alfv\'en waves in a model that resembles a solar coronal hole. Our main 
objective is to determine the role played by these waves in the spectral line 
broadening as observed in the coronal holes (e.g., Banerjee
et al. 1998; Dolla \& Solomon 2008).  We 
find an agreement between our numerical results of non-linear Alfv\'en waves 
and the computed line broadening of constituted synthetic spectra at different 
heights in model coronal hole and the observational data. This allows us to 
conclude that large-amplitude Alfv\'en waves are responsible for the observed 
non-thermal broadening of the spectral lines in the coronal holes.  Our result 
is important because it is an indirect evidence for the existence of non-linear 
Alfv\'en waves in solar coronal holes. 

The outline of the paper is as follows: our numerical model is described in
Sec. 2; a brief description of the used numerical code and the form of initial 
perturbations are given in Sec. 3; the results of our numerical simulations are 
presented in Sec. 4; comparison of our results to the observational data is 
given in Sec. 5; the obtained results are discussed in Sec. 6; and our 
conclusions are given in Sec. 7.

\section{Numerical  model of Alfv\'en waves}\label{sec:atm_model}
%
\subsection{MHD equations}\label{sec:equ_model}

Our model of the solar atmosphere contains a gravitationally-stratified 
magneto-plasma, which is described by the following set of ideal 
magnetohydrodynamic (MHD) equations:
\beqa
\label{eq:MHD_rho} 
{{\partial \varrho}\over {\partial t}}+\nabla \cdot (\varrho{\bf V})=0\, ,\\
\label{eq:MHD_V}
\varrho{{\partial {\bf V}}\over {\partial t}}+ \varrho\left ({\bf V}\cdot \nabla\right ){\bf V}= 
-\nabla p+ \frac{1}{\mu} (\nabla\times{\bf B})\times{\bf B} +\varrho{\bf g}\, , \\
\label{eq:MHD_B}
{{\partial {\bf B}}\over {\partial t}}= \nabla \times ({\bf V}\times {\bf B})\, , \\
\label{eq:MHD_divB}
\nabla\cdot{\bf B} = 0\, , \\
\label{eq:MHD_p}
{\partial p\over \partial t} + {\bf V}\cdot\nabla p = -\gamma p \nabla \cdot {\bf V}\, ,\\
\label{eq:MHD_CLAP}
p = \frac{k_{\rm B}}{m} \varrho T\, .
\eeqa
Here ${\varrho}$ is mass density, ${\bf V}$ and ${\bf B}$ are vectors of 
respectively the flow velocity and the magnetic field,
$p$ is gas pressure, $\gamma=5/3$ is the adiabatic index, ${\bf g}=(0,-g,0)$ is a vector of gravitational acceleration with 
its value $g=274$ m s$^{-2}$, $T$ is a temperature, $m$ is
a mean particle mass and $k_{\rm B}$ is a Boltzmann's constant.
\subsection {A model of the solar atmosphere}\label{sec:equil}

We consider a model of the solar atmosphere with an invariant coordinate 
($\partial/\partial z = 0$) and allow the $z$-components of velocity 
($V_{\rm z}$) and magnetic field ($B_{\rm z}$) to vary with $x$ and $y$.
In such 2.5D model, the solar atmosphere is in static equilibrium 
(${\bf V}_{\rm e}={\bf 0}$) with force- and current-free magnetic field, i.e.,
\beq\
(\nabla\times{\bf B}_{\rm e})\times{\bf B}_{\rm e} = {\bf 0}\, , \hspace{4mm} 
\nabla\times {\bf B}_{\rm e}={\bf 0}\, .
\label{eq:B}
\eeq
Henceforth the subscript $_{\rm e}$ corresponds to equilibrium quantities.

In our model of the atmosphere,
a curved magnetic field is given by
\beq
\label{eq:B_e}
{\bf B}_e = \nabla\times{\bf A_{\rm e}}\, ,
\eeq
where the magnetic flux function (${\bf A_{\rm e}}$) has the~form
\beq
\label{eq:A}
{\bf A_{\rm e}} = \Lambda_{\rm B} B_{\rm 0} \cos\left(\frac{x} {\Lambda_{\rm B}}\right)\exp{\left(-\frac{y-y_{\rm r}} {\Lambda_{\rm B}}\right)} {\bf\hat{z}}\, .
\eeq
Here
${\bf\hat{z}}$ is a unit vector along the $z$-direction and
$B_{\rm 0}$ is the magnetic field at the reference level, $y=y_{\rm r}$, 
that is chosen at $y_{\rm r}=10$ Mm. We set and hold fixed $B_{\rm 0}$ in such a way that 
the Alfv\'en speed, $c_{\rm A}=B_{\rm 0}/\sqrt{\mu \varrho_{\rm e}(y=y_{\rm r})}$ is ten times higher 
than the sound speed, $c_{\rm s}=\sqrt{\gamma p_{\rm e}(y=y_{\rm r})/\varrho_{\rm e}(y=y_{\rm r})}$. 
Such a choice of ${\bf B}_{\rm e}$ results in Eq.~(\ref{eq:B}) 
being satisfied.
Here $\Lambda_B=2L/\pi$ denotes the magnetic scale-height,
and $L$ is a half of the magnetic arcade width.
As we aim to model a polar region, we take $L=75$ Mm and keep it fixed in 
our calculations.  For such a choice, the magnetic field lines are weakly curved 
(not shown) and represent the open and expanding field lines similar to the 
coronal holes.

As a result of Eq.~(\ref{eq:B}), the pressure gradient is balanced by the gravity force
\begin{equation}\label{eq:p}
-\nabla p_{\rm e} + \varrho_{\rm e} {\bf g} = {\bf 0}\, .
\end{equation}
Using the ideal gas law of Eq. (\ref{eq:MHD_CLAP}) and the $y$-component of the hydrostatic
pressure balance indicated by Eq. (\ref{eq:p}), we express
equilibrium gas pressure and mass density as
\beqa\label{eq:pres} 
p_{\rm e}(y)=p_{\rm 0}~{\rm exp}\left( -
\int_{y_{\rm r}}^{y}\frac{dy^{'}}{\Lambda (y^{'})} \right),\hspace{4mm}
\varrho_{\rm e} (y)=\frac{p_{\rm e}(y)}{g \Lambda (y)}\, .
\eeqa
Here
\begin{equation} 
\Lambda(y) = \frac{k_{\rm B} T_{\rm e}(y)} {mg}
\end{equation}
is the pressure scale-height, and $p_{\rm 0}$ denotes the gas
pressure at the reference level. 

\begin{figure}
\begin{center}
\includegraphics[width=8.5cm]{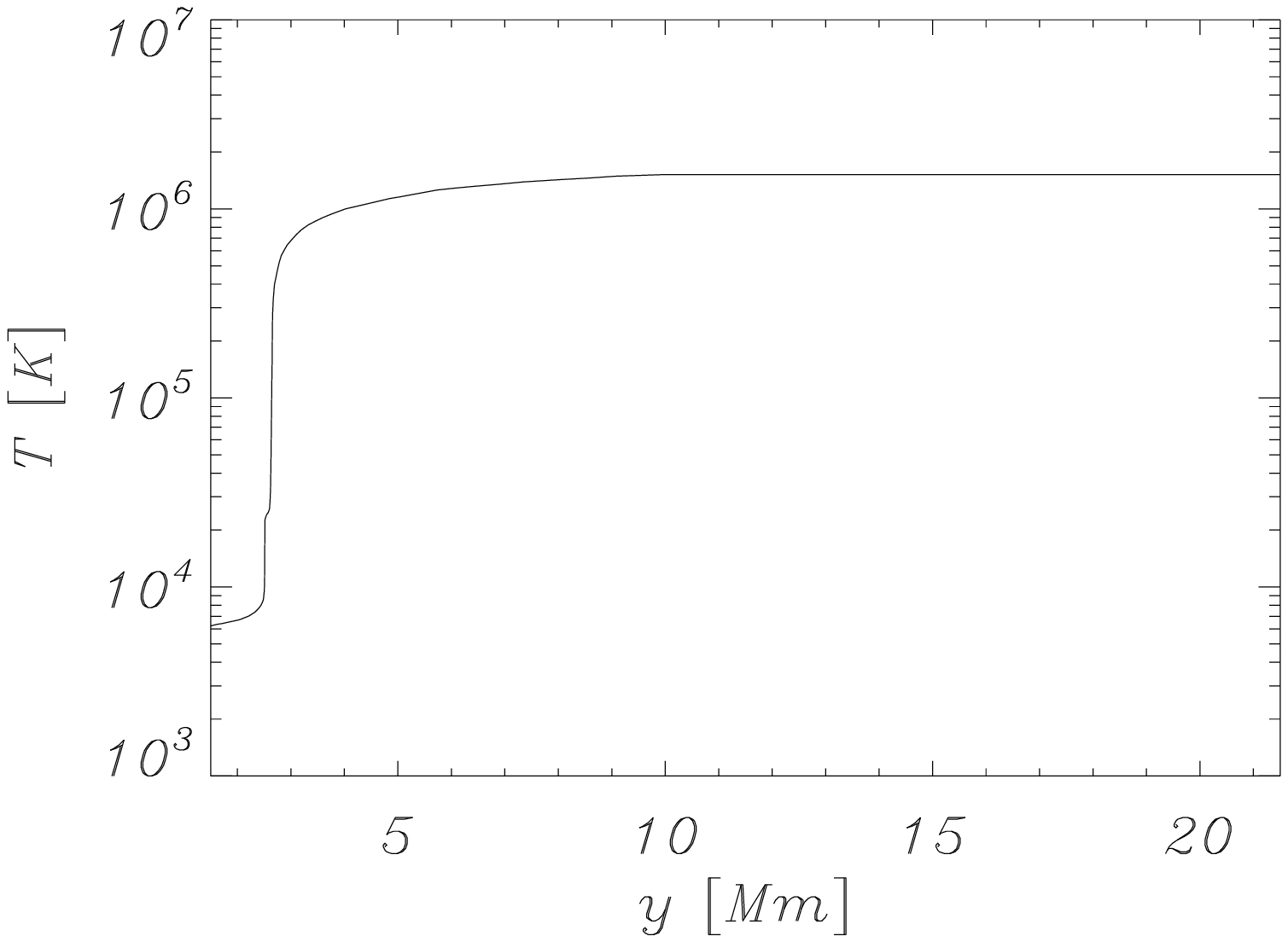}\\
\includegraphics[width=8.5cm]{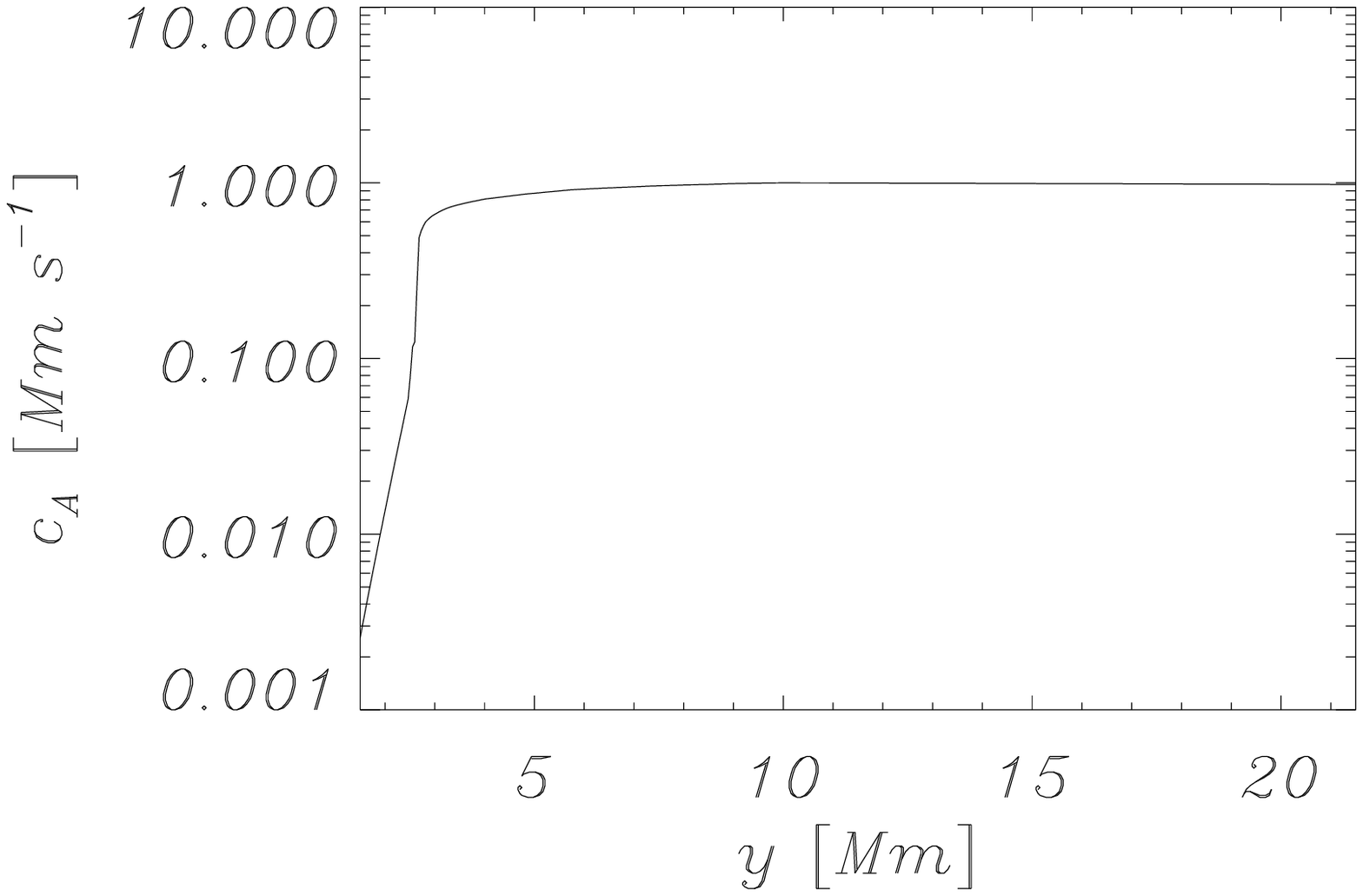}
\caption{\small 
Equilibrium profile of the temperature (top) and the Alfv\'en speed (bottom).
} 
\label{fig:Tealf}
\end{center}
\end{figure}

We adopt a realistic model of the plasma temperature profile (Vernazza~et~al.~1976),
displayed in Fig.~\ref{fig:Tealf} , top panel.
Temperature attains a value of about $6 \times 10^{3}$~K at $y=1.5$ Mm
and it grows up to about $1.5 \times 10^{6}$ K in the solar corona at $y=10$ Mm.  
Higher up temperature is assumed constant; note that this profile is more realistic 
than the one used by Murawski \& Musielak (2010).
The temperature profile determines uniquely the equilibrium mass density
and gas pressure profiles. 
Both $\varrho_{\rm e}(y)$ and $p_{\rm e}(y)$ experience a sudden drop
at the transition region that is located at $y \simeq 2.7$ Mm (not shown).
%
%
%

In this model the Alfv\'en speed, $c_{\rm A}$, varies only with $y$ and is expressed as follows: 
\beq
\label{eq:ca}
c_{\rm A}(y) = \frac{B_{\rm 0}e^{-\frac{y-y_{\rm r}}{\Lambda_{\rm B}}}}{\sqrt{\mu \varrho_{\rm e}(y)}}\, .
\eeq
Its profile 
is displayed in Fig.~\ref{fig:Tealf}, bottom panel.
Note that the Alfv\'en speed in the chromosphere, 
$c_{\rm A}(y=1.75\, {\rm Mm})$, is about $25$ km s$^{-1}$.
The Alfv\'en speed rises abruptly through the transition region reaching a value 
of $c_{\rm A}(y=10\, {\rm Mm}) = 10^{3}$ km s$^{-1}$ (Fig.~\ref{fig:Tealf}, bottom).
The growth of $c_{\rm A}(y)$ with height results from a faster decrement of $\varrho_{\rm e}(y)$
than  $B_{\rm e}(y)$ with the height.

The realistic solar atmosphere above the polar corona reveals  
complexity of its plasma and magnetic field structure.
The magnetic field configuration in the polar coronal hole can be approximated  
by expanding coronal funnels in the lower part of the atmosphere and comparatively 
smooth open field lines in its upper part (Banerjee et al. 1998; Hack-
enberg et al. 2000). This implies that
the field structure and magnetic scale-height vary above the polar corona. Moreover, 
the field configuration changes from dipolar to multipolar during the transition 
from the solar minimum to its maximum. Despite these well-known variations, 
we assume that the magnetic field scale-height is fixed at a resonable value 
of $L=75$ Mm in our simulation domain, which represents the weakly curved and open 
field lines of the polar coronal hole for the quiet minimum phase of the Sun. 
We want to point out that this assumption does not affect the validity of our 
numerical results.
%
%
\section{Numerical solutions of MHD equations}\label{sec:num_sim_MHD}
Equations (\ref{eq:MHD_rho})-(\ref{eq:MHD_CLAP}) are solved
numerically with a use of the FLASH code (Fryxell et al.
2000; Lee \& Deane 2009; Lee et al. 2009). This code
implements a second-order unsplit Godunov solver with various slope
limiters and Riemann solvers 
as well as Adaptive Mesh Refinement
(AMR) 
(MacNeice et al. 1999). We use the minmod slope limiter and the
Roe Riemann solver (e.g., Toro 2009). We set the simulation box as
$(-5\, {\rm Mm},5\, {\rm Mm}) \times (-1\, {\rm Mm},84\, {\rm Mm})$ 
and impose fixed in time boundary conditions for all plasma quantities
 in the $x$- and $y$-directions, while 
all plasma quantities remain invariant along the $z$-direction.
In our present work, we use AMR grid with a minimum (maximum) level of
refinement set to $3$ ($8$). The refinement strategy is based on
controlling the numerical errors in mass density. 
Blocks are denser below $3$ Mm and vertically along the region of Alfv\'en wave propagation.
Every numerical block consists of $8\times 8$ identical numerical cells. 
This results in an excellent resolution of vital spatial profiles and
greatly reduces the numerical diffusion at these locations.

%
\subsection{Initial perturbations}
We perturb initially (at $t=0$ s) the model equilibrium, described in Sec.~\ref{sec:equil}, 
by a Gaussian pulse in the $z$-component of velocity given by 
\beq
\label{eq:init_per}
V_{\rm z}(x,y,t=0) = A_{\rm v} \exp\left[ -\frac{(x-x_{\rm 0})^2+(y-y_{\rm 0})^2}{w^2} 
\right]\, ,
\eeq
where $A_{\rm v}$ is the amplitude of the pulse, $(x_{\rm 0},y_{\rm 0})$ is its 
initial position and $w$ denotes its width. 
We set $w=1$ ${\rm Mm}$, 
$(x_{\rm 0}=0, y_{\rm 0}=1.75)$ Mm and consider 
two cases: (a) $A_{\rm v}=25$ km~s$^{-1}$; (b) $A_{\rm v}=50$ km~s$^{-1}$.

Note that in  the 2.5D model, we developed, the Alfv\'en wave decouples from magnetoacoustic waves
and it can be described by $V_{\rm z}(x,y,t)$.
As a result, the initial pulse triggers Alfv\'en waves that in the linear limit are described by the wave equation
\beq
\label{eq:Vz_lin}
\frac{\partial^2 V_{\rm z}}{\partial t^2} = c_{\rm A}^2(y) \frac{\partial^2 V_{\rm z}}{\partial y^2}\, .\\
\eeq
%
%
\section{Results of numerical simulations}\label{sec:resultsOFnumSIM}

We simulate impulsively excited non-linear Alfv\'en waves and investigate
their propagation along the open magnetic field lines of a coronal hole model in 
the outward direction. 
It should be noted that the effect of the inhomogeneities 
across the magnetic field lines are not included in our approach. The waves are 
generated by the transversal velocity pulse perpendicular to the magnetic isosurface 
(X-Y) in the $z$-direction. 
This pulse is described by Eq.~(\ref{eq:init_per}). 

\begin{figure}
\centering
{
\includegraphics[width=8.25cm,angle=0]{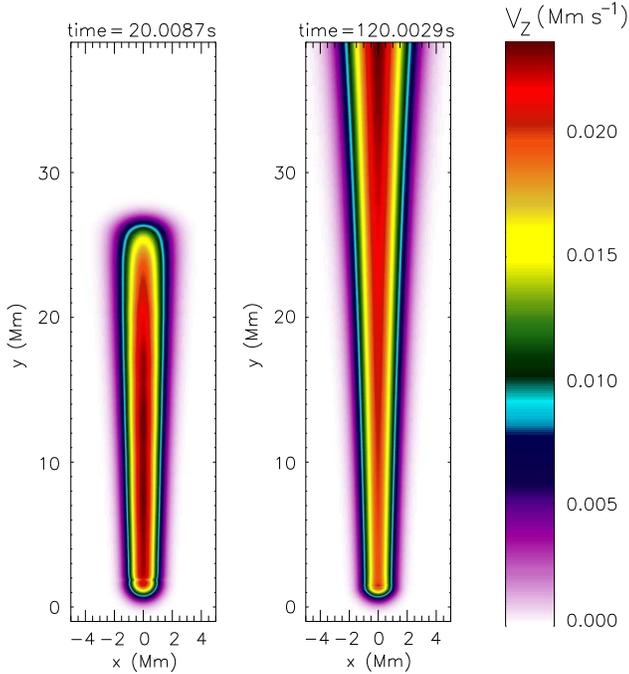}
}
\caption{\small
Transverse velocity $V_{\rm z}$ profiles 
at $t=20$ s 
and $t=120$ s for 
$A_{\rm v}=25$ km s$^{-1}$. 
}
\label{fig:6panels}
\end{figure}

\begin{figure*}
\begin{center}
\mbox{
\includegraphics[height=4.05cm]{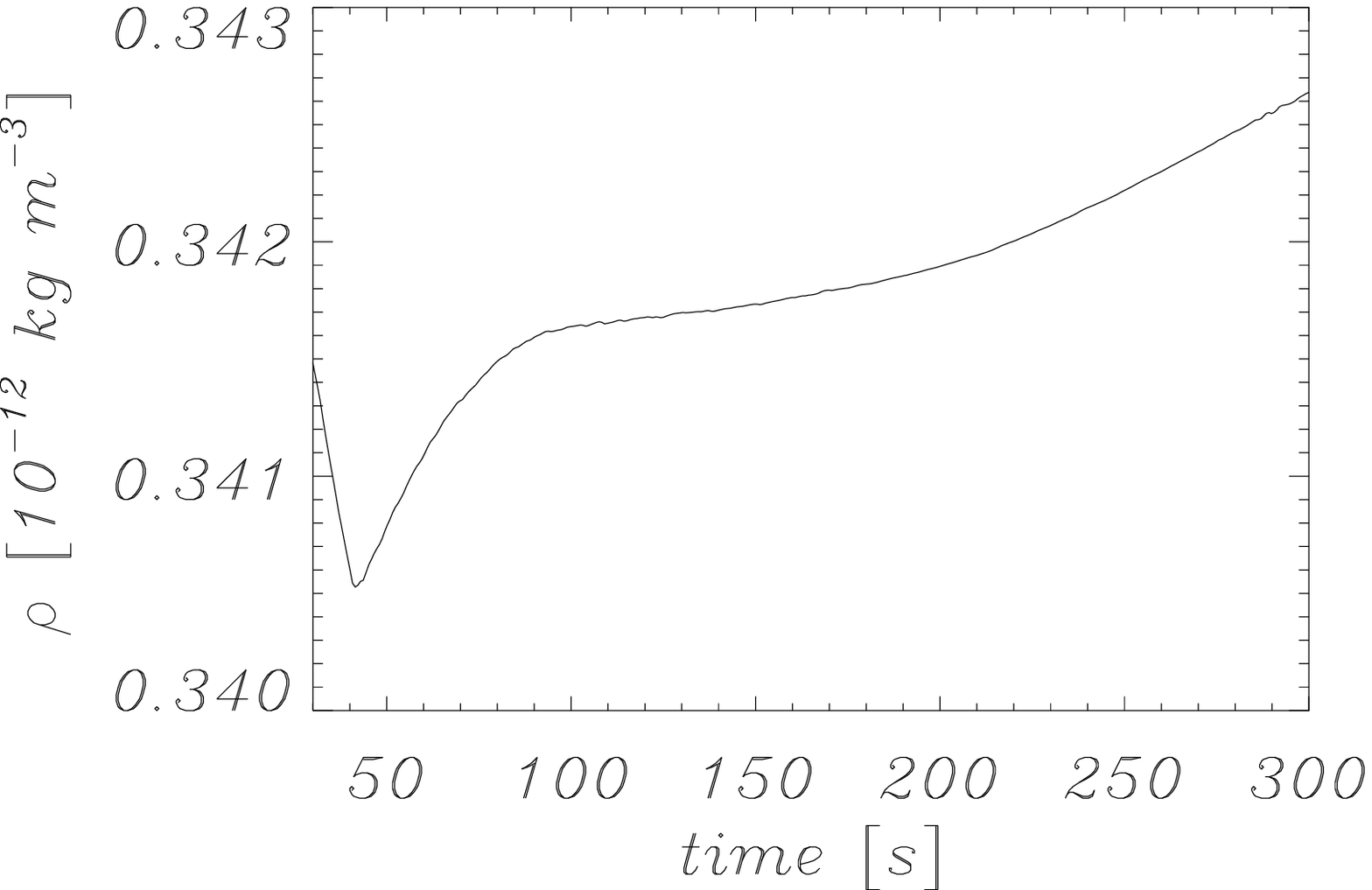}
\includegraphics[height=4.05cm]{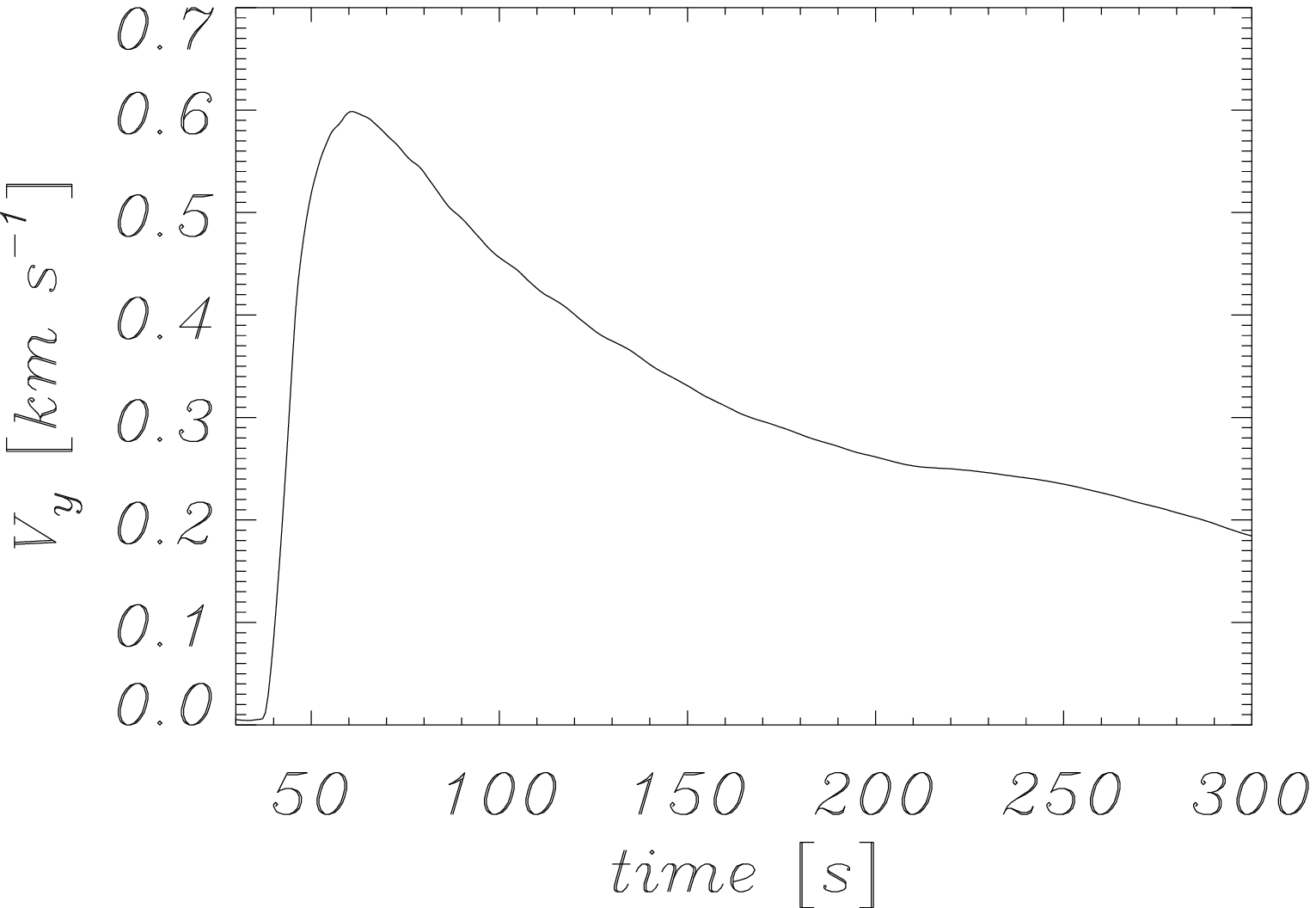}
\includegraphics[height=4.05cm]{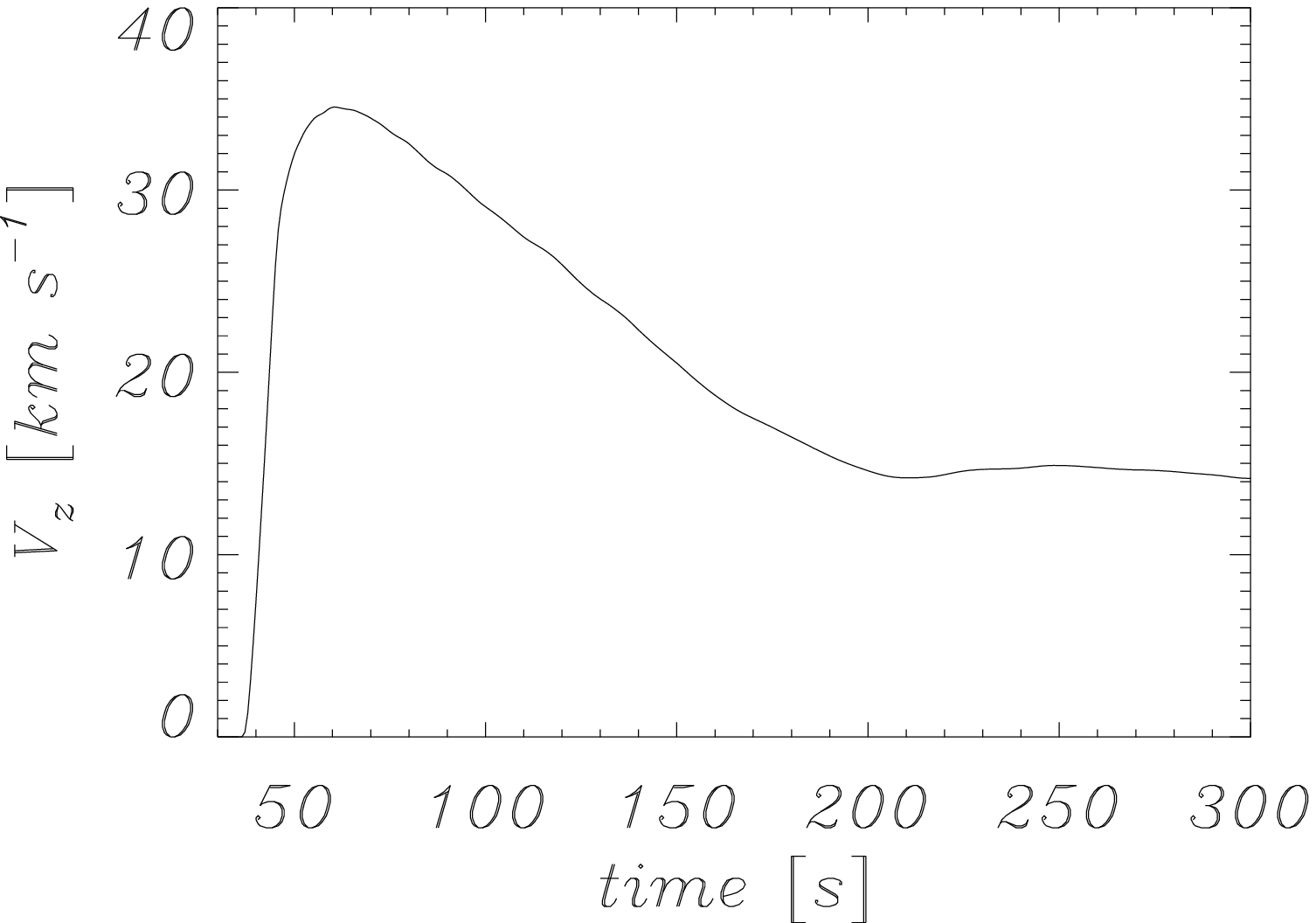}
}
\mbox{} 
\mbox{
\includegraphics[height=4.05cm]{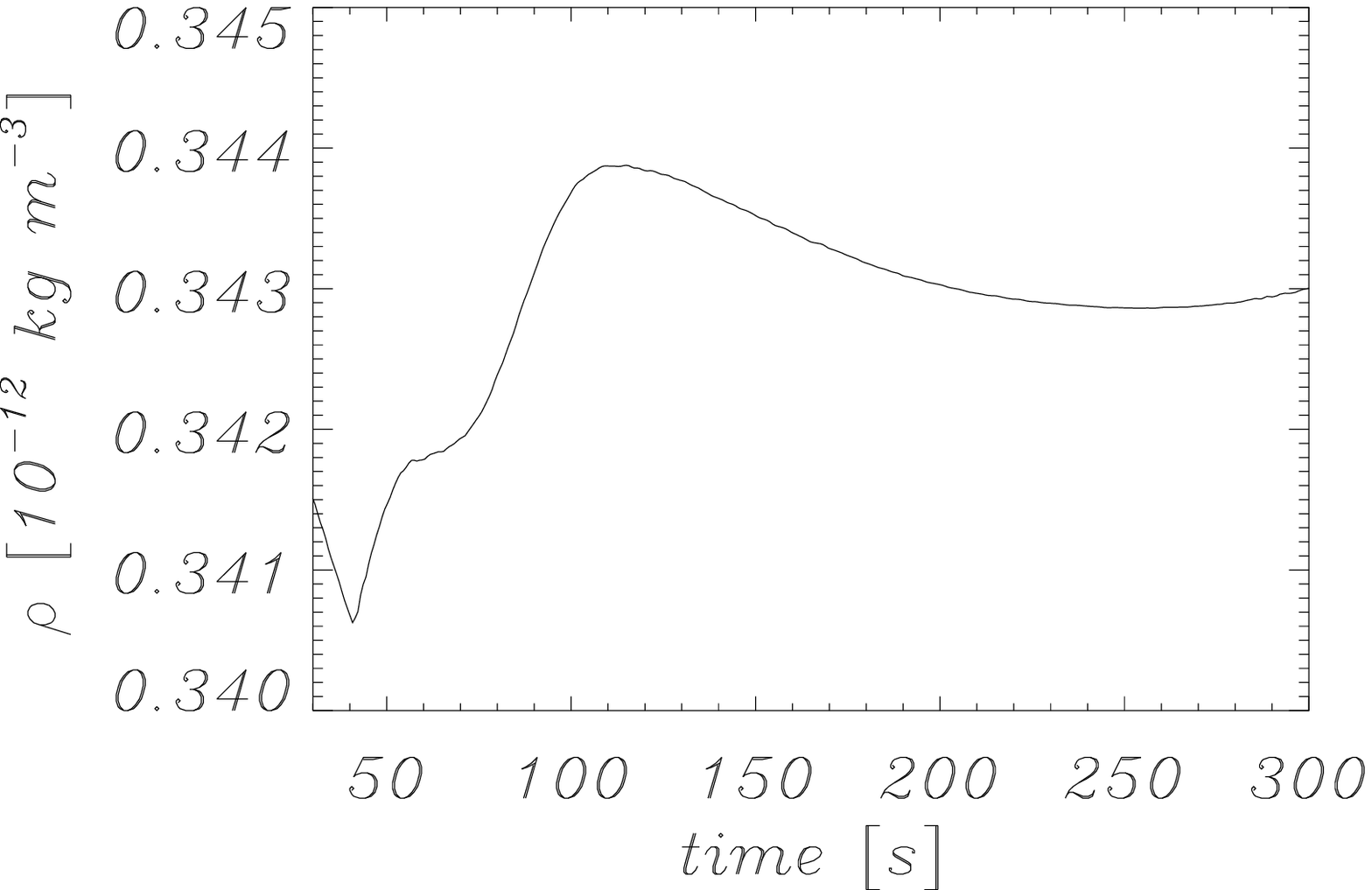}
\includegraphics[height=4.05cm]{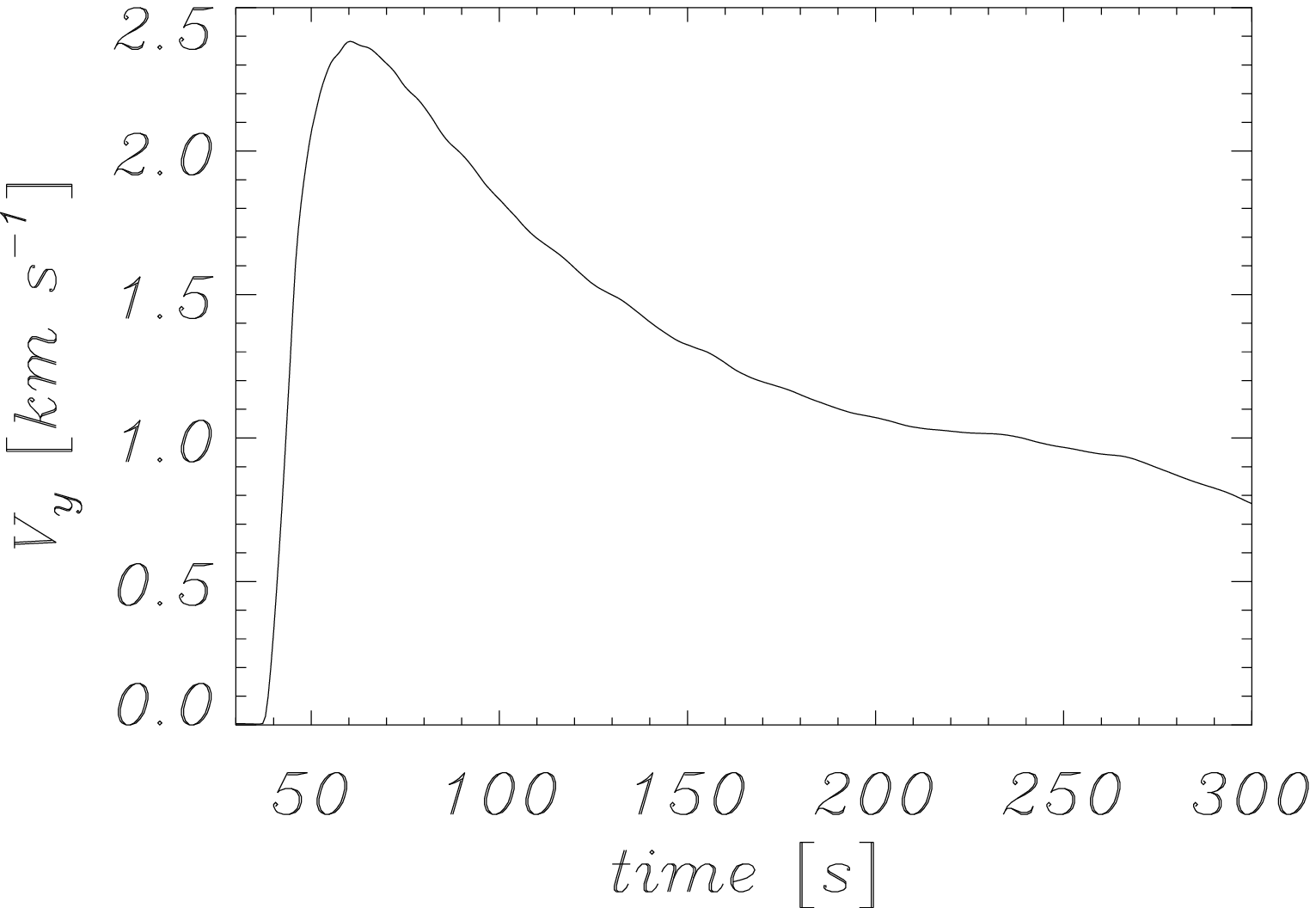}
\includegraphics[height=4.05cm]{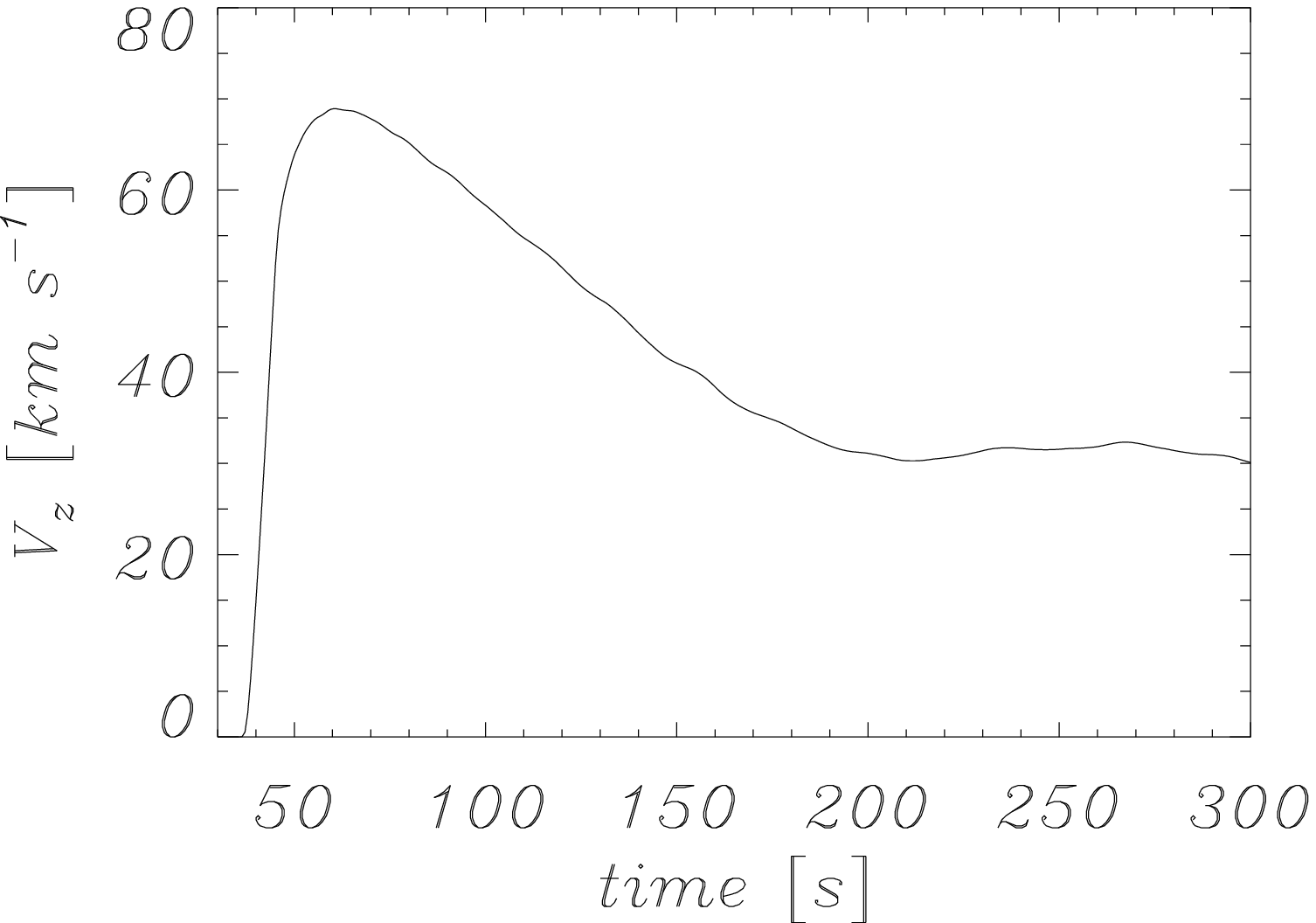}
}
\caption{\small
Results of the numerical
simulation of pulse-excited Alfv\'en waves: time-signatures of $\rho$, $V_{\rm y}$ and $V_{\rm z}$ for 
$A_{\rm v}=25$ km s$^{-1}$ (top panels) and
$A_{\rm v}=50$ km s$^{-1}$ (bottom panels) 
collected at $(x=0, y=49)$ Mm.
} 

\label{fig:vz}
\end{center}
\end{figure*}   
%

First, we consider the pulse amplitude equal to $25$ km~s$^{-1}$. 
Spatial profiles of $V_{\rm z}$ for $A_{\rm v}$=25 km s$^{-1}$ and the corresponding time signatures obtained 
in our numerical simulations of Alfv\'en waves are shown respectively 
in Figs.~\ref{fig:6panels} and \ref{fig:vz}, top-right panel. 
Perturbations of $V_{\rm z}$ propagate essentially along magnetic field lines,
which is displayed on spatial profiles of $V_{\rm z}$ at $t=20$~s and $120$~s (Fig.~\ref{fig:6panels}).
%
A part of the whole simulation region is illustrated.
Note that the Alfv\'en wave experiences an acceleration at that height
as $c_A(y)$ increases up to $1$ Mm s$^{-1}$ at the transition region 
(Fig.~\ref{fig:Tealf}, right).
This acceleration of Alfv\'en wave is caused by significant fall in mass density
at essentially constant magnetic field. 
Above, the Alfv\'en wave propagates with slightly increased amplitude and almost constant 
velocity. The spatial resolution of the transition region is set to be $\Delta x = \Delta y = 0.04$ Mm in the simulation domain.

We also simulate the Alfv\'en wave using the large transversal pulse of 
$A_{\rm v}=50$ km~s$^{-1}$ 
that generates the large-amplitude, 
non-linear Alfv\'en waves.
Time-signatures of V$_{z}$ for $A_{\rm v}=50$ km~s$^{-1}$ collected at ($x = 0$, $y = 49$) Mm
are shown in right-bottom panel of Fig.~\ref{fig:vz}.
Since the wave is non-linear, it generates a vertical flow (Fig.~\ref{fig:vz}, middle-bottom panel) and 
mass density perturbations (Fig.~\ref{fig:vz}, left-bottom panel), which 
are driven by the ponderomotive force.

We collect the temporally and spatially averaged transversal velocity component V$_{z}$ at each height of the simulation 
domain (y$_{max}$ = 84 Mm) generated by the pulses of different wave amplitudes, 
$A_{\rm v}$= 25 and 50 km s$^{-1}$ (see Fig. 3).
If we assume that the $y$-direction of the simulation 
box is placed along the outward open magnetic field of a polar coronal hole, then at 
each height the transversal amplitude of Alfv\'enic perturbations may contribute to 
the otherwise unresolved non-thermal motions by affecting the line-width of the observed 
spectral lines. Finite-amplitude Alfv\'en waves propagating along the open magnetic field lines of the 
polar coronal holes can perturb the plasma velocity, which causes positive and negative 
Doppler shifts that can be detected as a line width broadening or line width variation 
(Banerjee et al. 1998; Harrison et al. 2002; Moran 2003; O'Shea et
al. 2005; Dwivedi \& Srivastava 2008). We now compare our results of numerical 
simulations with the available observational data. 

\section{Comparison to observational data}

\subsection{Line width broadening}

The non-thermal motion is a prominent candidate that may modify the observed line 
profiles  by its contribution to the full width at half maximum (FWHM) of the 
spectral lines. Banerjee et al. (1998) have estimated the radial variation of the non-thermal
velocity in the polar coronal hole by deducing its contribution to the observed line 
profiles of Si VIII $\lambda$1445 \AA\ . The sophisticated line-width measurements 
and thus the observed variation of non-thermal velocity are very few in the available 
literature. Using SOHO/CDS observations, O'Shea et al. (2005) have shown the line width increment 
in the inner corona and then its decrement  beyond radial distance 1.21 R$_{o}$ in coronal 
hole.  Here R$_{o}$ is the radius of the Sun.  However, due to the unavailability of the instrumental width of SOHO/CDS slits, the exact informations about the non-thermal velocity 
could not be derived, so the estimate gave only the uncorrected line-width variation with 
the height. Moreover, Banerjee et al. (1998) have approximately re-produced the trend of the spatial variation of non-thermal velocity in the polar coronal hole similar to Banerjee et al. (1998) using 
2$"$ slit of Hinode/EIS. Note that the exact instrumental width of EIS slits were variables 
with wavelength of the observed spectral lines. 

By using the temperature averaged over the simulation domain, the averaged density 
generated from the simulations, and the temporally and spatially averaged transversal 
speed (V$_{\rm z}$) of Alfv\'en wave generated by the velocity pulse $A_{\rm v}$ = 50 
km s$^{-1}$, we have estimated the equivalent FWHM ($\sigma$) of its 1445.75 \AA\ line at 
different heights using the relation (Mariska 1992)
\beq
\label{eq:FWHM}
\sigma^{2} = \left[ 4\ln2\left(\frac{\lambda}{c}\right)^2 \left( \frac{2k_{\rm B}T}{m_{\rm i}}+\xi^2 \right)+{\sigma_{I}^{2}} \right]\, .
\eeq

It should be noted that the averaged wave velocity amplitude can be scaled appropriately 
in terms of non-thermal speed as $\xi^2=0.5\,{V^{2}}_{z}$ by taking into consideration 
of polarization and direction of the propagation of wave w.r.t line-of-sight (Banerjee et al. 1998).  
We adopt the same scaling throughout the paper, although different authors use different 
scalings based on their assumption of the degree of freedom of wave motion (Dolla \& Solomon 2008).  
At any particular height in the model coronal hole, we average the wave velocity amplitude 
of transversal oscillation in temporal domain between $t_{\rm a}$-250 to $t_{\rm a}$+250 s, 
while in the spatial domain over the entire pulse width, where $t_{\rm a}$ is the arrival 
time of a wave signal to the detection point.  Therefore, the resultant transversal motion contributes to the non-thermal unresolved motion at each height in the model coronal hole 
plasma as observed in form of spectral line broadening by various authors, e.g., Banerjee et al. (1998) and Dolla \&
Solomon (2008).

\subsection{Line profiles obtained with $A_{\rm v} =$ 50 km s$^{-1}$}

We computed the synthetic line profiles (cf., Fig.~4) of Si VIII $\lambda$1445.75 \AA\ at 
three different heights $y=$ 20, 40, and 70 Mm based on the inclusion of the non-thermal 
contribution of Alfv\'en wave excited by a pulse with its amplitude $A_{\rm v} =$ 50 km s$^{-1}$. The CHIANTI atomic data base is used to produce the synthetic line profiles by computing the  
averaged density ($\rho$), temperature (T), and FWHM estimated by averaged velocity (V$_{z}$). 
The CHIANTI SSW routine $"$synthetic.pro$"$ is used to produce the line profiles; note that the routine takes into account the ionic equilibrium as reported by Mazzotta et al. (1998), the coronal hole differential emission measure (DEM) values, and coronal abundances as available in CHIANTI 
atomic data base. Although, the major input parameters are the averaged density ($\rho$) (or pressure), temperature (T), and FWHM (see Eq. \ref{eq:FWHM}) derived from the simulation, while other parameters mentioned above are considered as optional input parameters in the calculation 
of synthetic spectrum.

In the present case, 
we compare the deduced theoretical line width with the corrected observed one as reported by Banerjee et al. (1998) using SoHO/SUMER. Therefore, 
we neglected the implication of instrumental width of SoHO/SUMER slits in Eq.~(16).  We synthesized the line 
profiles of the Si VIII 1445.75 \AA\ line but did not calculate the continuum as we wanted 
to only show the shape of the line profiles at various heights and their broadening that 
compares well with the observations of Banerjee et al. (1998).  The line width of synthetic line 
profiles of Si VIII 1445.75 \AA\ at 20 Mm, 40 Mm, and 70 Mm (Eq.~16) are respectively 353 m\AA\ , 358 m\AA\ , and 363 m\AA\ when we consider the non-thermal contribution of impulsively excited Alfv\'en wave by a pulse with its amplitude $A_{\rm v}$=50 km s$^{-1}$.  It is clear from Table 2 of Banerjee et al. (1998)
that the estimated corrected line-widths of Si VIII $\lambda$1445.75 \AA\ at heights 27 arcsecs ($\sim$19.6 Mm), 57 arcsecs ($\sim$ 41.3 Mm), 98 arcsecs ($\sim$71 Mm) are respectively $\sim$292 m\AA\ , $\sim$334\AA\ , and $\sim$369 m\AA\ .

We also adopted and tested a different way of calculating the line-profiles at a particular 
height (e.g., 15 Mm) in the model coronal hole.  Using the results of our numerical simulation, 
we derived the plasma parameters ($\rho$, T) as well as the transversal speed V$_{z}$ for each instant set ($X$, $Y$=fixed, $t$) in which X-space is averaged over 1.0 Mm spatial scale.  The time instant was considered with each 10 s resolution between $t$-250 s to $t$+250 s.  We collected the parameters at each instant, and then derived the averaged FWHM, which was found to be 284 m\AA\ for a pulse $A_{\rm v}$=50 km s$^{-1}$.   The result closely matches the broadened line width of 291 m\AA\ as reported by Banerjee et al. (1998) at a height of 23 arcsec. Therefore, the alternative estimations are also consistent with our findings.

Although our theoretically estimated line-width also shows increment with height in our model coronal hole, the spatial gradient of increment is rather flat compared to the observations by Banerjee et al. (1998).  The most likely reason is that we considered the impulsive excitation of the pulse-driven Alfv\'en wave that also contributes more to the un-resolved non-thermal motion 
near the wave source region in the lower solar atmosphere. Therefore, the computed line-width, 
,for instance at 27 arcsecs ($\sim$19.6 Mm), is higher than the observed line width given by Banerjee et al. (1998) at the 
same height. However, the theoretically estimated and the observed line-widths at coronal heights e.g., at 40 and 70 Mm, closely match each other.

The results presented in Fig. 4 clearly show that the synthetic line profiles computed with the wave amplitude $A_{\rm v} =$ 50 km s$^{-1}$ closely approximately mimic the observed spectrum of Si VIII $\lambda$1445.75 \AA\ obtained by the detector-B on SOHO/SUMER and reported by Banerjee et al. (1998).  This provides an
evidence that the large-amplitude, pulse-excited Alfv\'en wave can contribute upto some extent into the broadening of the spectral lines observed by Banerjee et al. (1998).

\subsection{Line profiles obtained with $A_{\rm v} =$ 25 km s$^{-1}$}

We again derived the synthetic line profiles (cf., Fig.~5) of Si VIII $\lambda$1445.75 \AA\ (Dolla \&
Solomon 2008) at three different heights $y=$ 43.5, 65.25, and 72.5 Mm, based on the inclusion 
of the non-thermal contribution of Alfv\'en wave excited by the pulse with its amplitude $A_{\rm v} = 25$ km s$^{-1}$.  The CHIANTI atomic data base and the method described above was again used to produce the synthetic line profiles.  The obtained results show that there is not very significant broadening of line profiles in our model coronal hole for these smaller amplitude Alfv\'en waves; for the observational data see Dolla \&
Solomon (2008).

The line widths of Si VIII $\lambda$1445.75 \AA\ synthetic line profiles at 43.5 Mm ($\sim$60$"$), 65.25 Mm ($\sim$90$"$), and 72.5 Mm ($\sim$100$"$) (Fig.~5) above the limb in our coronal hole model are estimated respectively as $\sim$289 m\AA\, $\sim$290 m\AA\, and 291 m\AA\ when we consider the non-thermal contribution of impulsively excited Alfv\'en wave by a pulse strength of $A_{\rm v}$=25 km s$^{-1}$. Dolla \&
Solomon (2008) have reported the Gaussian half line-widths of various coronal spectral lines (e.g., Si VIII $\lambda$1445.75 \AA\, Fe XII $\lambda$1242, Mg X $\lambda$624 \AA\ ) between $\sim$42 Mm (57$"$) and $\sim$100 Mm (140$"$) in the polar corona hole (cf., their Fig.~8).  This shows that our theoretical estimations are below the observationally estimated line-width as reported by Dolla \&
Solomon (2008) for Si VIII $\lambda$1445.75 \AA\ between 40-75 Mm in coronal hole.  For example, the estimated observed half line-widths at 1/$\sqrt(e)$ as reported by Dolla \&
Solomon (2008) (cf., their Fig 8), after conversion in FWHM of Si VIII $\lambda$1445.75 \AA\ at heights $\sim$43.5 Mm ($\sim$60$"$), $\sim$ 65.25 Mm ($\sim$90$"$), and $\sim$72.5 Mm ($\sim$100$"$), respectively give $\sim$362 m\AA\, $\sim$368\AA\ , and $\sim$376 m\AA\ .

Our theoretically estimated line-width shows very small increment w.r.t. height in the coronal hole model between 40-75 Mm height.  Now, the observed width of Si VIII 1445.75 \AA\ also varies very little but with comparatively higher gradient than theoretical values.  The most probable reason is that the plasma and magnetic field properties of the observed coronal hole differs from the assumed conditions in our model, in which the Alfv\'en wave of smaller amplitude propagates without much growth as well as its contribution in the line broadening.

Our numerical results obtained for non-linear Alfv\'en waves with the amplitude $A_{\rm v}$=25 km s$^{-1}$ show that the resulting contribution of these waves to non-thermal motions approximately qualitatively match the observational data of spectral line width as observed by Dolla \&
Solomon (2008) for Si VIII $\lambda$1445 \AA\ line of SoHO/SUMER, as both show the lesser spectral line broadening.  However, the theoretically estimated widths are comparatively smaller than the observed ones at various heights in coronal hole.  Although, the theoretically derived line-widths only approximately match the observed ones at various heights 
in the coronal hole, there is no clear evidence of significant line broadening in these cases.
%
%
\begin{figure*}
\begin{center}
\mbox{
\includegraphics[width=4.5cm, angle=90]{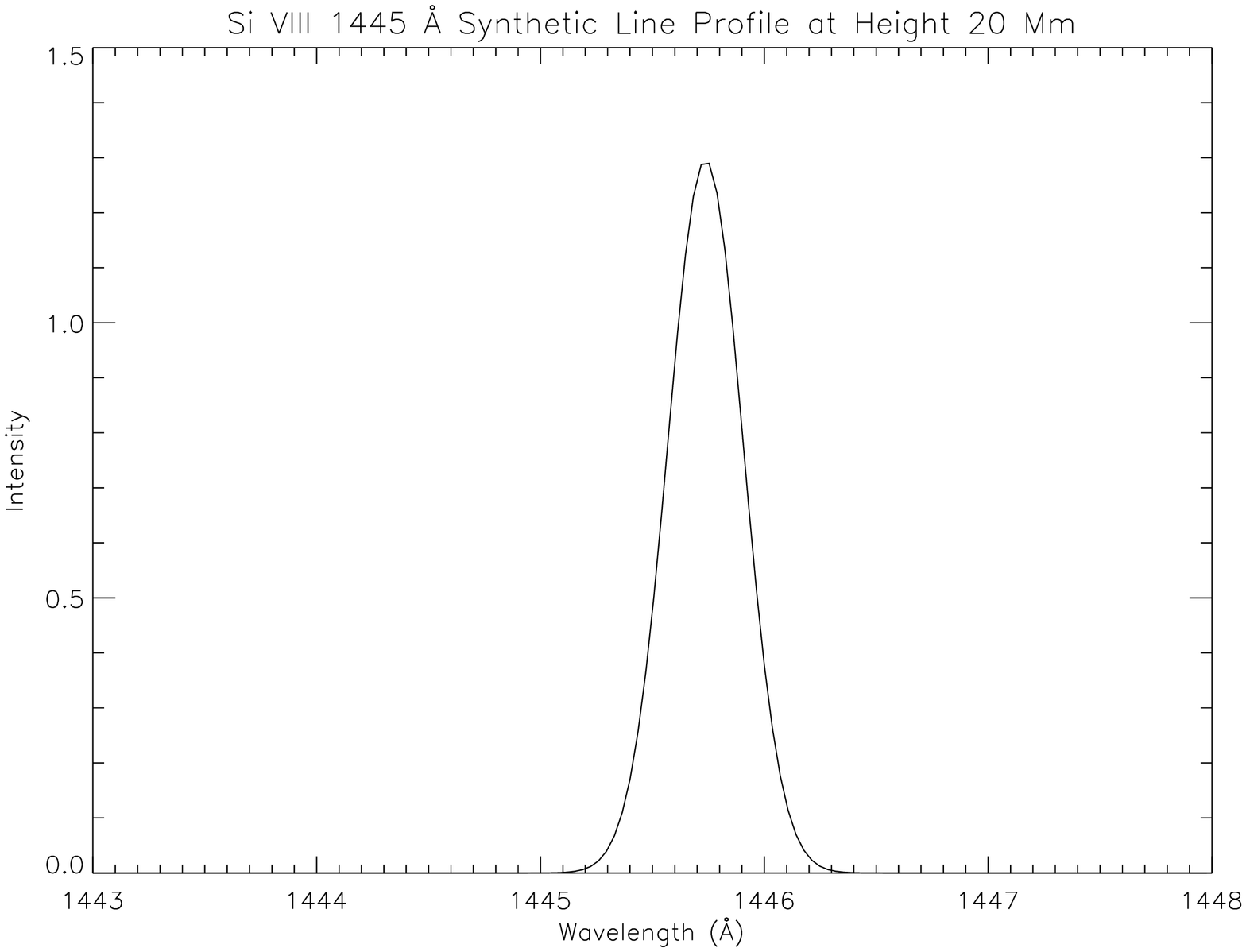}
\includegraphics[width=4.5cm, angle=90]{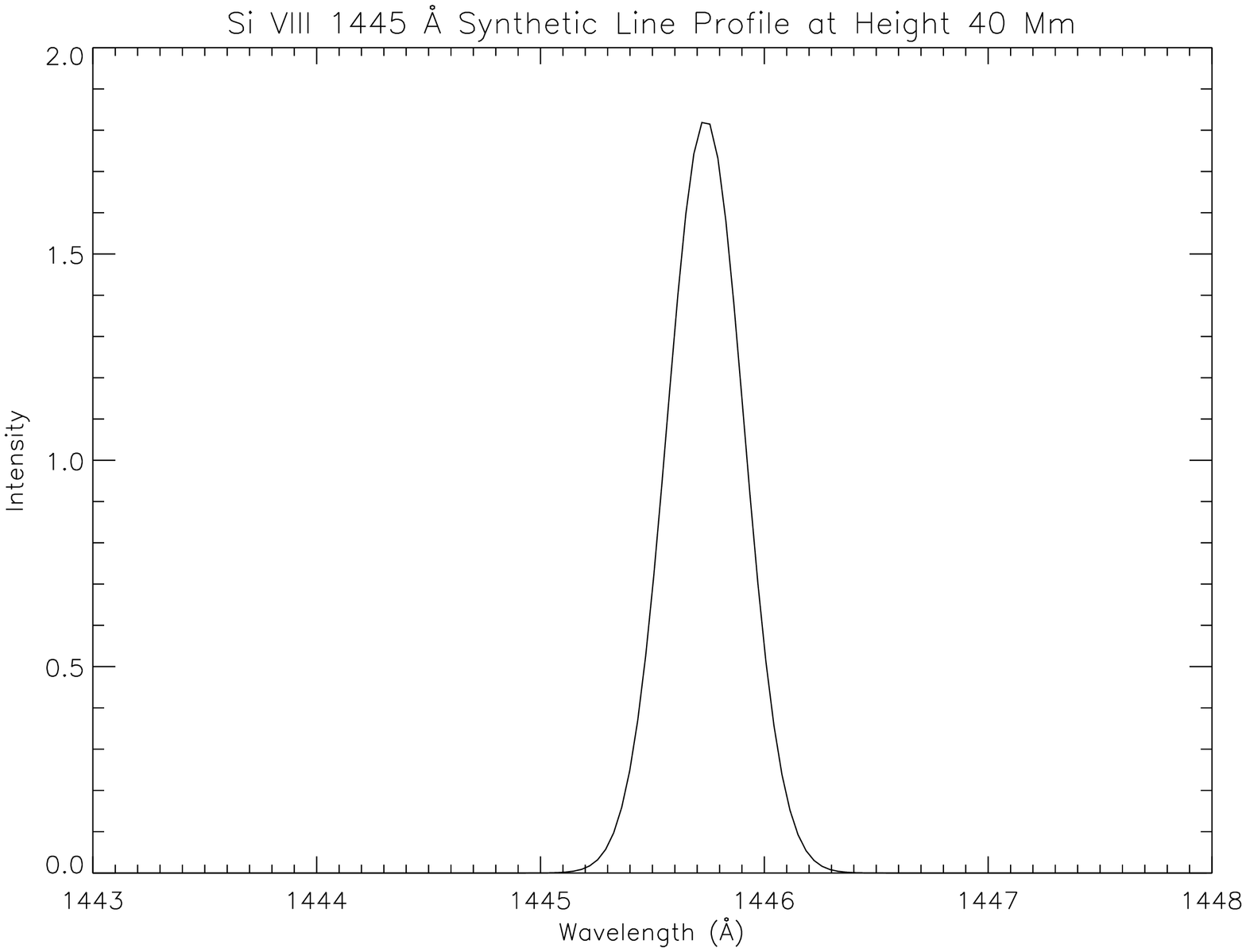}
\includegraphics[width=4.5cm, angle=90]{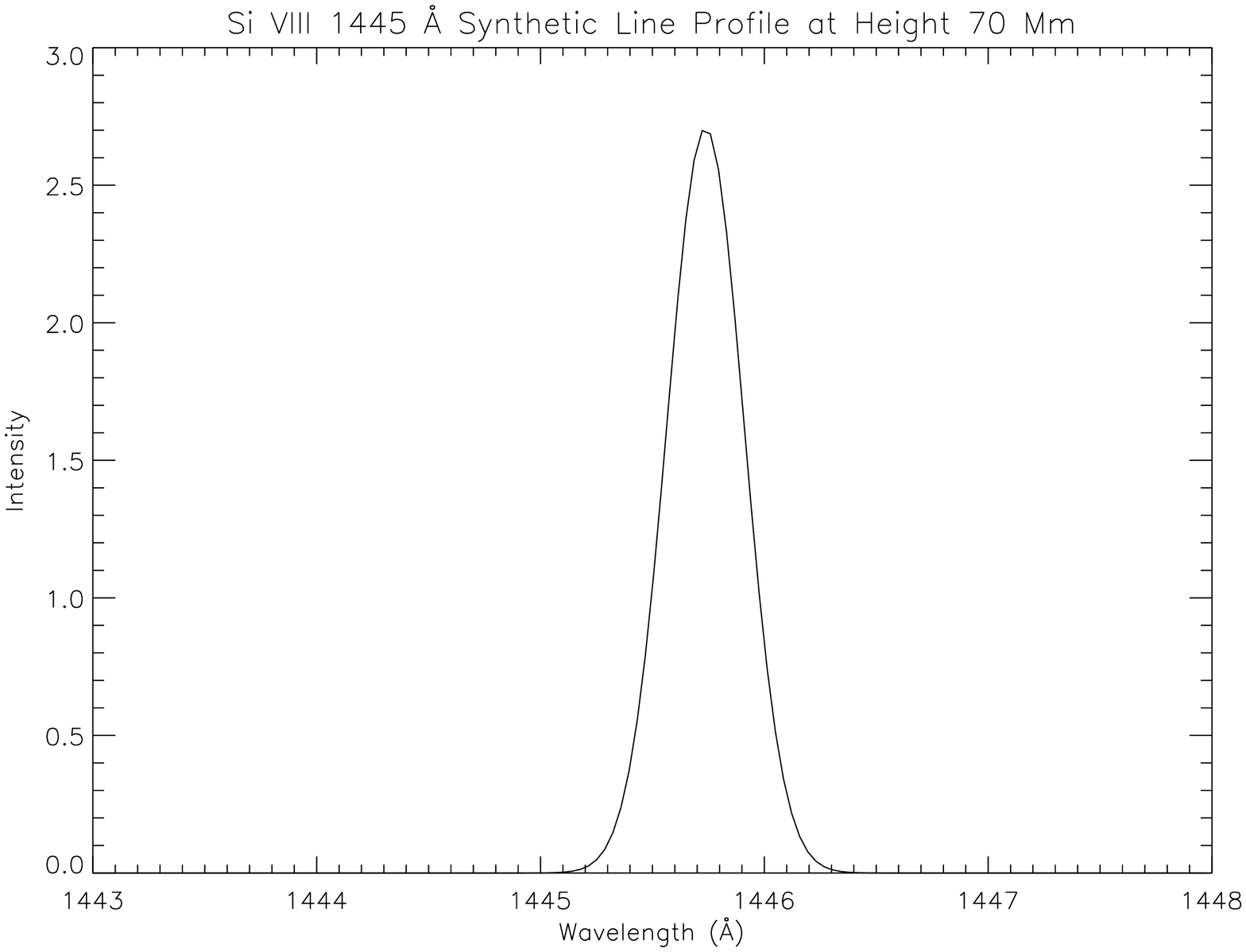}}
\caption{\small The synthetic line profiles of Si VIII 1445.75 \AA\ line
at three different heights $y=$ 20, 40, and 70 Mm to compare with the 
observations of Banerjee et al. (1998), based on the inclusion of the non-thermal 
contribution of Alfv\'en wave excited by pulse amplitude $A_{\rm v} =$ 50 km s$^{-1}$.
The line-profile is deduced by considering the ionic equilibrium as reported by 
Mazzotta et al. (1998), the coronal hole DEM values, and coronal abundances 
as available in CHIANTI atomic data base. The simulated
temperature and density are also used as 
input parameters.}

\label{fig:synt}
\end{center}
\end{figure*}

\section{Discussion}   
As already mentioned in Sec. 1, there are several analytical models that 
attempted to explain the observed non-thermal broadening of spectral lines 
in the solar corona by using linear Alfv\'en waves (e.g., Harrison et al. 2002
; Pek{\"u}nl{\"u} et al. 2002; Dwivedi \& Srivastava 2006).  
Under the linear theory of propagating Alfv\'en waves
from the solar photosphere upward into the solar corona along the magnetic 
field lines, it is found that the non-thermal velocity is inversely 
proportional to the quadratic root of the observationally estimated electron 
density that matches well the characteristics of such waves 
(e.g., Banerjee et al. 1998; Moran et al. 2003; Baner-
jee et al. 2009).  The falls-off of the density with height 
can amplify the propagating Alfv\'en wave and increase its group velocity
that can contribute in the rms wave amplitude and thus in the non-thermal 
spectral line-width broadening with height 
in the solar atmosphere.  

The most relevant numerical work has been done by Kudoh \& Shibata (1999)
, who used numerical simulations to investigate the role of torsional Alfv\'en waves 
in both the formation of spicules as well as the observed non-thermal 
broadening of spectral lines in the solar corona.  They considered Alfv\'en 
waves to be generated by perturbations of 1 km s$^{-1}$ in the solar photosphere. 
According to these authors, the waves can lift the spicules up to 5000 km, 
provide the energy flux 3.0 $\times$ 10$^{5}$ ergs s$^{-1}$ cm$^{-2}$, and 
produce the non-thermal broadening of emission lines in the solar corona 
with approximately 20 km s$^{-1}$.  

It is clear from the above discussion that neither previous analytical 
methods nor previous numerical simulations have been able to explain the 
full profile of the observed spectral line-width variation
with height in the polar coronal holes. Nevertheless, it is 
noteworthy here that previous observational results have been matched 
with linear propagating waves coming from photosphere (Banerjee et al. 1998,
2009). 
They have showed, in linear regime, that the rms wave amplitude is related 
with some powers of mass density and magnetic field variations.  In this paper, 
we are not attempting this indirect analytical model explanations but instead 
we directly compute the temporally and spatially averaged
transversal velocity component, $V_{\rm z}$, from 
our pulse-excited Alfv\'en wave model and determine its contribution to rms 
wave velocity amplitude. It should be noted then that the non-thermal
velocity motion caused by such waves is inherent in the 
computed line-width variation of synthetic Si VIII 1445.75 \AA\ line
with height in the model coronal hole, and that it is consistent 
with the observations reported by Banerjee et al. (1998)
. Hence, we conclude 
that the computed spectral line 
broadening with height in the coronal hole implies the increment
of non-thermal velocity as provided by the Alfv\'en waves.
The results of our numerical 
simulations of non-linear Alfv\'en waves show that the non-thermal 
spectral line width broadening in polar coronal holes 
can be explained by these waves.  

In our pulse-excited model, the pulse of a large amplitude is launched 
above the solar photosphere and this pulse splits into the upward and 
backward propagating wave-trains in the overlying solar atmosphere. 
The upward moving pulse train causes the instantaneous displacement of 
the field lines in perpendicular plane away and towards the line-of-sight.  
This effect generates the transversal velocity and thus the Doppler shift 
as well as line broadening.  Therefore, in our case of pulse-excited 
non-linear Alfv\'en wave model, the more stringent contribution of 
averaged transversal velocity V$_{z}$ component at a particular 
height causes the non-thermal broadening of the spectral line there. 

The theoretically estimated 
line-width of Si VIII 1445.75 \AA\ synthetic line increased with height 
in the model coronal hole due to the pulse-excited, large amplitude Alfv\'en waves,
which is consistent with the observations of the same phenomenon 
in the polar coronal hole (Banerjee et al. 1998). We have also generated synthetic line profiles
of Si VIII 1445.75 \AA\ line (Fig.~4) 
that match well the sample observations of Banerjee et al. (1998)
showing the spectral
line broadening. These measurements may also be the 
signature of the undamped and growing transversal pulse train that is the 
pulse-excited Alfv\'en wave.  However, the region of our interest
is below 80 Mm in the solar atmosphere, where the observations
show the line-width increment consistent with the 
excitation of the undamped Alfv\'en waves.
It should be noted that the region of decreasing line-width
as reported by O'Shea et al. (2005) is quite high that was interpreted 
in terms of Alfv\'en wave dissipation.
We do not consider such regime of observations in our model, which instead is focused 
on the Alfv\'enic dynamics in the lower solar atmosphere of the coronal 
hole that is also the source region of the fast solar wind. 

It is well known that the fast solar wind 
starts accelerating near 20 Mm in the magnetic funnels in the polar 
coronal regions (Tu et al. 2005). Therefore, the undamped and growing pulse-excited, 
large amplitude Alfv\'en waves can provide the momentum to the 
solar wind plasma in the inner corona like the classical Alfv\'en waves,
which are typically considered to be generated in the solar photosphere.  
On the other hand, the generation of large-amplitude Alfv\'enic pulses 
above the solar surface considered in this paper is still a puzzle.  
The reconnection events between the emerging magnetic fields with the 
existing open magnetic field lines may be one of the most plausible 
mechanisms that can trigger such strong transversal velocity pulses 
and thus drive Alfv\'en waves along the open field lines of the coronal 
hole. However, such waves and their physical properties may be highly 
dependent on the local magnetic field configuration of the reconnection 
region (Kigure et al. 2010). 

Recently, Okamoto \& De Pontieu (2011)
have used the high-resolution observations 
obtained by the Hinode instruments and discovered the 
propagating Alfv\'en waves along solar spicules with a velocity
amplitude of $\sim$7 km s$^{-1}$. In addition, McIntosh et al. (2011)
have found evidence for the outward-propagating Alfv\'enic 
motions with amplitudes of the order of 20 km s$^{-1}$
and periods of the order of 100-500 s throughout the quiescent 
atmosphere, and concluded that these motions carry enough energy 
to accelerate the fast solar wind and heat the quiet corona. These 
discoveries of the Alfv\'enic motions 
imply that the carried wave energy may cascade through various 
layers of the solar atmosphere. The main aim of this paper is 
to understand the role of the pulse-excited Alfv\'en waves in the 
unresolved non-thermal motion in polar coronal hole
and their role in the observed non-thermal spectral line broadening. 

In summary, the recent observations clearly show Alfv\'enic disturbances with amplitudes smaller 
than $50$ km s$^{-1}$ at spatial and temporal scales achieved by the simulations. 
This is a potentially limiting factor for the realism of the simulations. 
\begin{figure*}
\begin{center}
\mbox{
\includegraphics[width=4.5cm, angle=90]{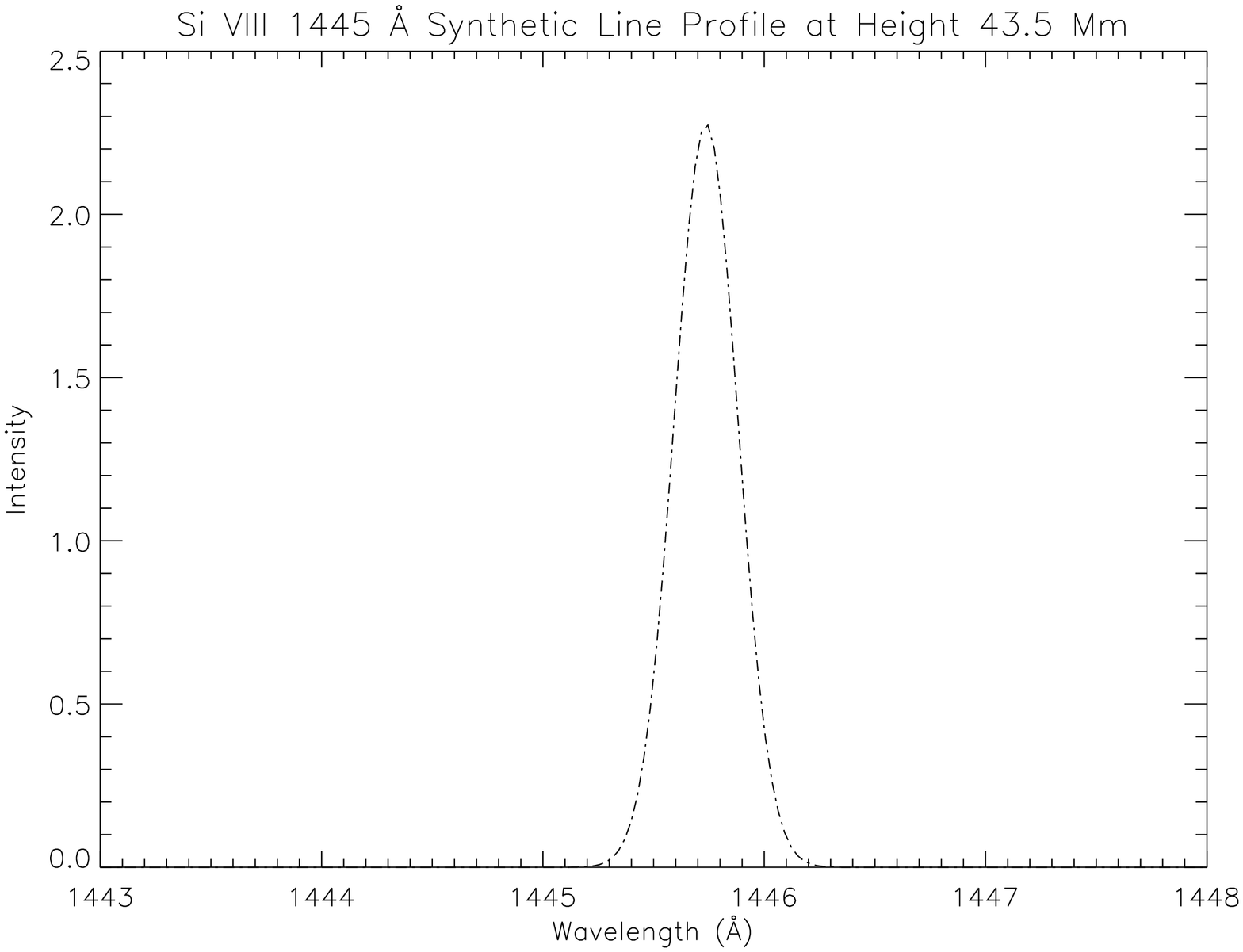}
\includegraphics[width=4.5cm, angle=90]{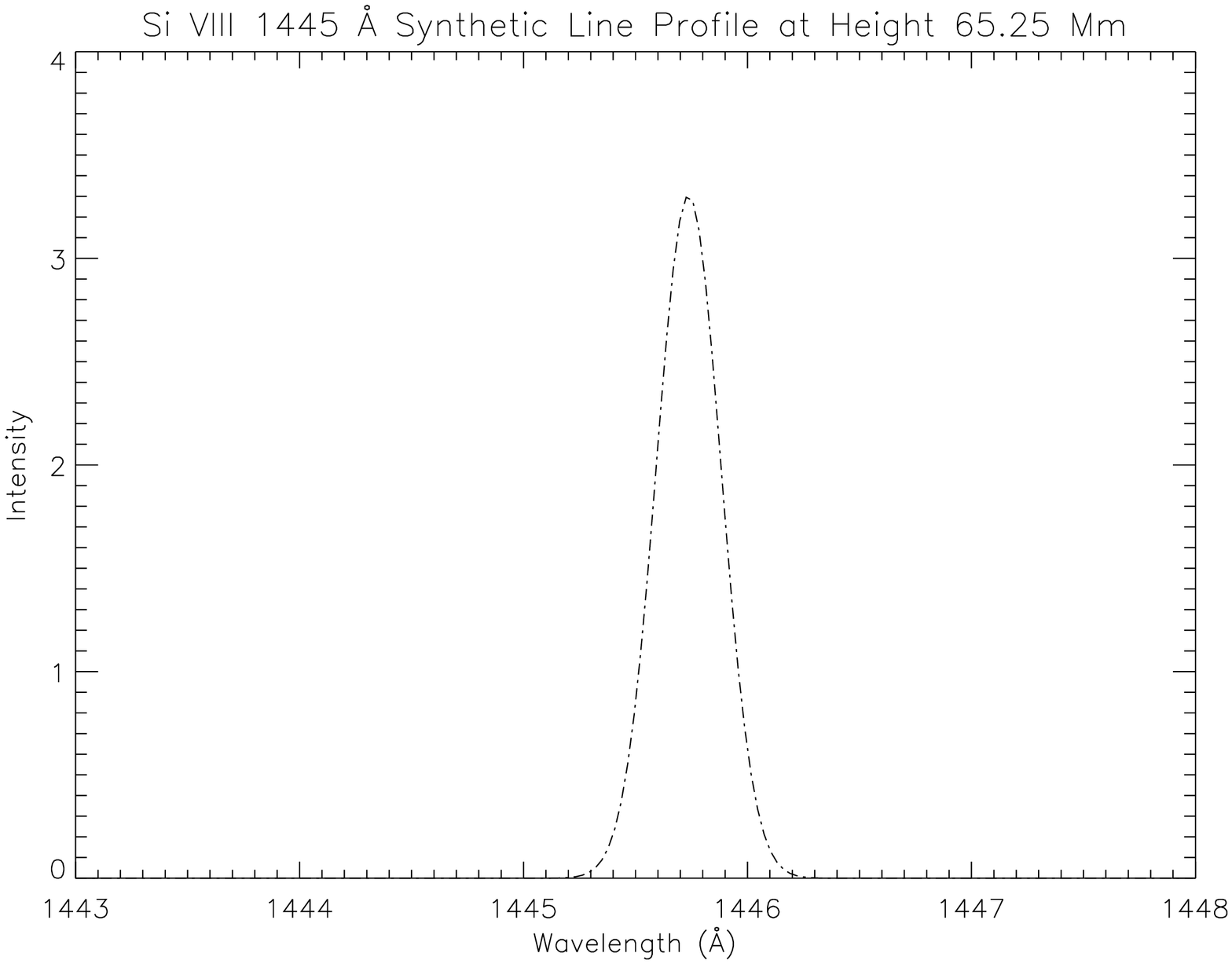}
\includegraphics[width=4.5cm, angle=90]{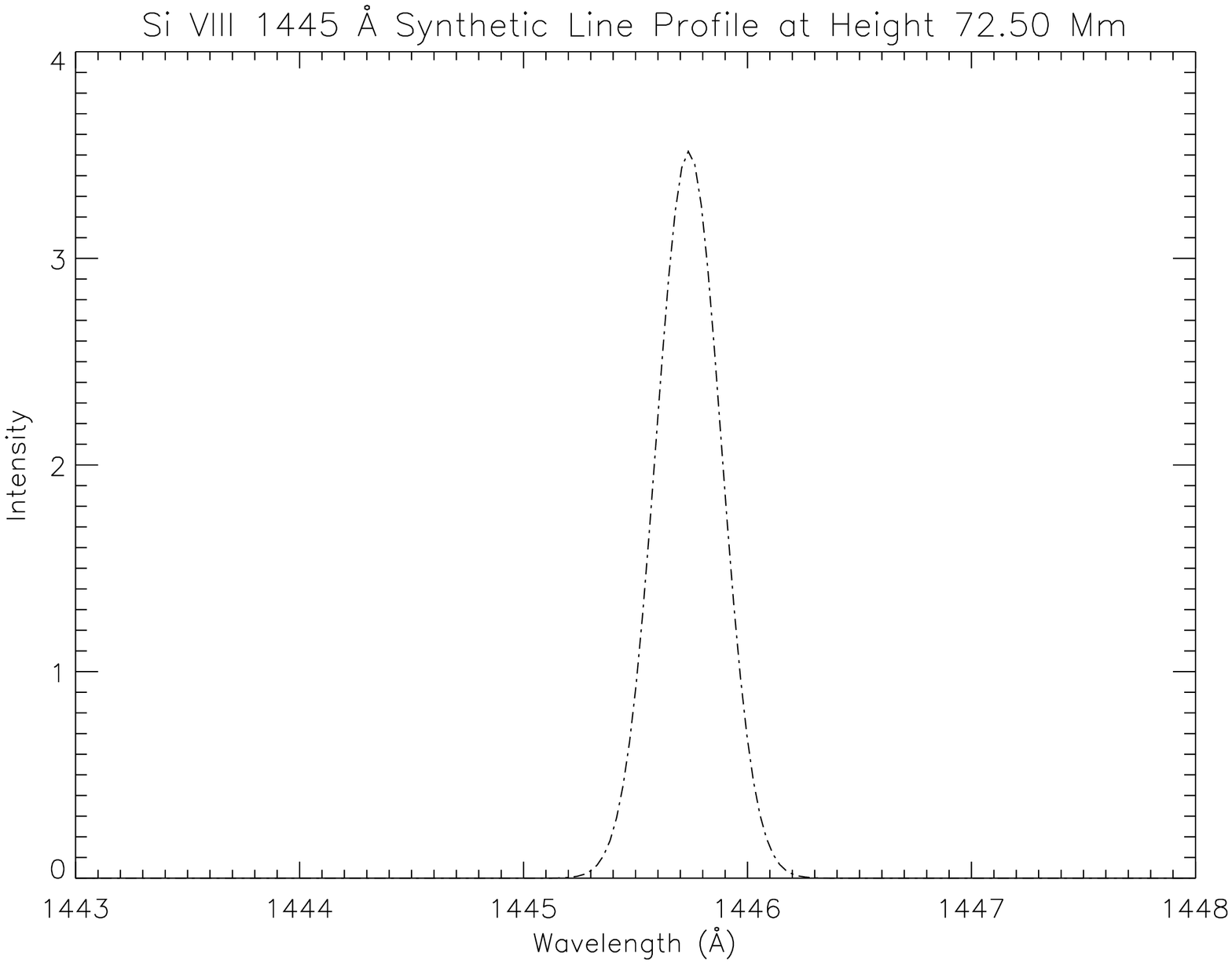}
}
\caption{\small The synthetic line profiles of SI VIII $\lambda$1445.75 \AA\ line
at three different heights $y=$43.5, 63.25, and 79.5 Mm in model coronal hole, based 
on the inclusion of the non-thermal contribution of Alfv\'en wave excited by pulse 
amplitude $A_{\rm v} =$ 25 km s$^{-1}$. The line-profile is deduced by considering 
the ionic equilibrium as reported by Mazzotta et al. (1998), the coronal hole 
DEM values, and coronal abundances as available in CHIANTI atomic data base. 
The simulated temperature and density are also used as input parameters.
These synthetic line profiles does not show any significant 
spectral line broadening as observed by Dolla \& Solomon (2008).}
\label{fig:synt}
\end{center}
\end{figure*}

The energy carried by Alfv\'en waves has been considered to be 
an important candidate for the heating of the coronal hole and the
acceleration of the solar wind by many authors (e.g., Ofman \& Davila 1995, 1998; Ofman 2005, and references
cited there). The dissipation of the energy carried
by these waves occurs either in the distant part of the corona (Parker 1991), 
or by some unique processes, such as phase-mixing, taking place in 
in the solar atmosphere (Heyvaerts \& Priest
1983; Hood et al. 1997; Nakariakov et al. 1997). The large-amplitude 
non-linear Alfv\'en waves can also be dissipated by coupling to 
longitudinal wave motions in the outer part of the magnetized solar 
corona (e.g., Boynton \& Torkelsson
1996). This shows that the solar atmosphere 
above the polar corona may be the ideal place for the growth of linear 
and non-linear Alfv\'en waves without much damping, and that the waves 
may carry their energy to the outer part of the corona and can heat the 
solar wind ions (e.g., Ofman \&
Davila 2001).

The main objective of the present study was to compute the averaged
amplitude of the pulse-excited Alfv\'en waves at different atmospheric
heights, and explore its effects on the broadening of the observed 
spectral lines.  Our computational domain covered both the lower part 
of the solar atmosphere as well as the inner solar corona, and we did
not take into account any dissipative processes in our numerical 
simulations because it was suggested that such processes may contribute 
to the observed narrowing of the line profiles (O'Shea et al. 2005). However, 
this work is out of the scope of this paper and will be considered as 
a future project.

\section{Conclusions}

In this paper, we simulated numerically the behavior of impulsively 
excited non-linear Alfv\'en waves in solar coronal holes, and studied for 
the first time their role in the observed broadening 
of spectral lines caused 
by these waves.  We compared the obtained numerical results to the 
spectral line broadenings observed by Banerjee et al. (1998);
Dolla \& Solomon (2008). 
We found that the large-amplitude, non-linear, pulse-driven ($A_{\rm v}$=50 km s$^{-1}$) Alfv\'en waves 
are the most likely candidates for the non-thermal
broadening of Si VIII $\lambda$1445.75 \AA\ line profiles in the polar coronal hole (Banerjee et al. 1998). 
Our results also show that Alfv\'en waves driven by comparatively 
smaller velocity pulse, $A_{\rm v}$=25 km s$^{-1}$, only approximately contribute to the observed 
spectral line width of Si VIII $\lambda$1445 \AA\ at various heights 
in coronal hole as reported by Dolla \& Solomon (2008), however, without significant 
evidence of broadening. 
The results of our numerical simulation and their comparison to the observations of
line broadening are important as they become an indirect evidence for the existence 
of larger amplitude, pulse-driven Alfv\'en waves in solar coronal holes. 

Finally, we would like to suggest that more spectroscopic 
observations should be carried out in future using  
high-resolution observations (e.g., Hinode/EIS, and also with upcoming 
Solar-C instruments) to search for a direct evidence of
such pulse-excited non-linear Alfv\'en waves in the solar atmosphere.
Obviously, more detailed theoretical studies of a variety of pulse-driven 
non-linear Alfv\'en waves excited by a range of pulse amplitudes are also
needed in order to better understand the role played by these waves in
the observed line broadening and the solar wind acceleration.
%
%
%

{\bf Acknowledgments.}
The software used in this 
work was in part developed by the DOE-supported ASC/Alliance Center for 
Astrophysical Thermonuclear Flashes at the University of Chicago. This 
work has been supported by the Alexander von Humboldt Foundation (Z.E.M.) 
and 
by a Marie Curie International Research
Staff Exchange Scheme Fellowship within the 7th European Community
Framework Program (P.Ch. \& K.M.) as well as by 
the "HPC Infrastructure for Grand Challenges of Science and Engineering" 
Project, co-financed by the European Regional Development Fund under 
the Innovative Economy Operational Program 
(P.Ch. \& K.M.).
We also acknowledge the CHIANTI,
which is a collaborative project involving researchers at NRL 
(USA) RAL (UK), and the Universities of Cambridge (UK), 
George Mason (USA), and Florence (Italy).
A.K.S. thanks Shobhna Srivastava for patient encouragement, 
and also to Prof. D. Tripathi, IUCAA, Pune, India
for valuable discussions on spectral line-profiles 
and their implications.
%



\end{document}